\documentclass{elsart}

\usepackage{natbib}

\usepackage{epsfig}

\usepackage{amssymb}

\begin{document}

\begin{frontmatter}



\title{Turbulence, Complexity, and Solar Flares}


\author{R. T. James McAteer\corauthref{cor}\thanksref{footnote1}}
\address{School of Physics, Trinity College Dublin, Dublin 2, Ireland}
\corauth[cor]{Corresponding author}
\thanks[footnote1]{Marie Curie Fellow}
\ead{james.mcateer@tcd.ie}
\ead[url]{http://grian.phy.tcd.ie/$\sim$mcateer}

\author{Peter T. Gallagher, Paul A. Conlon}
\address{School of Physics, Trinity College Dublin, Dublin 2, Ireland}

\begin{abstract}

The issue of predicting solar flares is one of the most fundamental in physics, addressing issues of plasma physics, high-energy physics, and modelling of complex systems. It also poses societal consequences, with our ever-increasing need for accurate space weather forecasts. Solar flares arise naturally as a competition between an input (flux emergence and rearrangement) in the photosphere and an output (electrical current build up and resistive dissipation) in the corona. Although initially localised, this redistribution affects neighbouring regions and an avalanche occurs resulting in large scale eruptions of plasma, particles, and magnetic field. As flares are powered from the stressed field rooted in the photosphere, a study of the photospheric magnetic complexity can be used to both predict activity and understand the physics of the magnetic field. The magnetic energy spectrum and multifractal spectrum are highlighted as two possible approaches to this. 

\end{abstract}


\end{frontmatter}

\parindent=0.5 cm


\section{Introduction}

Solar flares are among the most energetic events in the solar system and have intrigued generations of physicists.  These events influence a panorama of physical systems, from the photosphere of the Sun, through the heliosphere and into geospace. Flares occur in active regions in the solar corona, volumes in the solar atmospheric plasma characterised by increased emission at 1,000,000 K or more and containing kilogauss magnetic fields. Active regions are believed to be formed through the convective action of subsurface fluid motions pushing magnetic flux tubes through the photosphere. These active region flux tubes are jostled around by turbulent photospheric and sub-photospheric motions and, when conditions are right, the active region produces a flare.  The energy of these events (typically  $\approx10^{32}$~erg) is thought to come from the energy stored in the magnetic field being suddenly converted into other forms, accelerating particles to near-relativistic speeds, and creating temperatures in excess of 10,000,000 K.  The precise conditions required to create these enormously energetic events are as yet unknown.
	
Solar flares are one of the main aspects of the larger field of space weather. Space weather generally refers to the interaction of solar particles and magnetic fields with the Earth's magnetosphere and upper atmosphere. Understanding this interaction is of considerable practical importance because technological systems, such as communications and navigation satellites, can suffer interruptions or permanent damage. Energy releases as extreme as solar flares and coronal mass ejections (CMEs), both in terms of their physics and potential human impact, are of fundamental importance in any consideration of space weather. This has long been recognized, from their first observation (R. C. Carrington, September 1st, 1859, recorded a white light flare), to the inception of sunspot classification systems (Hale 1919; McIntosh 1990), to current space weather considerations (Baker et al. 2008). In this paper we concentrate solely on the solar flare aspect of space weather prediction.

The concept of forecasting space weather is dependent on our ability to predict parameters of solar flares in advance using remote sensing data. The most important aspects of any solar flare are size ( a proxy for energy released in one flare), frequency ( a proxy for total energy released over some timespan), onset time, and location. An accurate onset time and location prediction would allow scientists to predict the geo-effectiveness of any event. Realistically a full accurate prediction is impossible with remote sensing data. However we can hope to obtain a probability measure of flare occurrence (at least we could predict an 'all-clear') by studying the magnetic field of active regions. Giovanelli (1939) showed that size, type (i.e., magnetic configuration) and development of an active region could be used to predict flare activity. K\"{u}nzel (1960) recognised the importance of the photospheric magnetic field measurements in the concept of adding the $\delta$ configuration into the Mount Wilson classification. The presence of a $\delta$ configuration, where large values of opposite polarity exist close together, was identified as a warning of the build up of magnetic energy stress, and hence the possibility of the occurrence of a large solar flare. Mayfield \& Lawrence (1985) showed a good correlation between flare energy and magnetic energy and found a $\delta$ component doubled the probability of a large flare. Sammis, Tang \& Zirin (2000) showed that large regions classified as containing a $\delta$ had a 40\% chance of producing a large event of size X1 or greater. These studies used this qualitative concept of magnetic complexity to show a maximum flare size could be estimated over some future period of time. An alternative `Zurich' system (Keipenheuer, 1953), modified by McIntosh (1990), has also shown some flare predictive qualities (Bornmann \& Shaw 1994) who show that even for the 60 different possible classes, flare predictions are of limited use for space weather purposes. 

There are two main approaches to the flare prediction problem. One is to combine existing active regions classifications systems and historical records. This approach is adopted by Sammis, Tang \& Zirin (2000) to study the historical maximum flare size from the Mound Wilson classes. Gallagher et al. (2002) make a statistical estimate of flare occurrence above some threshold, based on a historical record of McIntosh classes. Related to this, Bayesian studies (Wheatland 2001, 2004) of the waiting time distributions seem to do well at predicting the frequency of large events, although may over predict for more moderate size flares. Qahwaji \& Colak (2007) and Colak \& Qahwaji (2008) overcomes the qualitative nature of this approach somewhat by using machine learning to classify each active region and assign a flare prediction. Many of these approaches currently produce flare predictions in near-realtime which are made available online (www.swpc.noaa.gov, www.solarmonitor.org, spaceweather.inf.brad.ac.uk). 

The second technique is to try to quantitatively assign an indicator of magnetic energy or complexity to active regions and test the evolution of this against the flares from the region. Abramenko et al. (2002, 2003) and Abramenko (2005b) suggest the scaling index as a useful single parameter measure of magnetic complexity. McAteer, Gallagher \& Ireland (2005a) suggest the fractal dimension seems to present a minimum threshold for flare size. Schrijver (2007) suggest that an `R-value', a weighted measure of flux near a neutral line, scales with solar flare size. Falconer, Moore, Gary (2002, 2003, 2006) propose a number of different measures which scale with increased helicity and non-potentiality, and show how these can be used to predict CME productivity. The role of helicity is also studied by Nindos \& Andrews (2004) who connect the amount of helicity to whether a flare will be eruptive or confined. Leka \& Barnes (2003) adapt a large number of measures of magnetic complexity and show, in a statistical sense, which measures may act as best predictors of flare occurrence. These studies are quantitative and mostly automated, so could conceivably provide near-realtime predictions. However the current lack of full-disc vector magnetic field information limits the usefulness of many of them for space weather prediction.

In this paper we present the basic physics confirming the usefulness of using magnetic field information in solar flare prediction. We concentrate on characterizing active region magnetic field complexity in an attempt to begin to understand which active region properties are important indicators of their activity.  In Section~\ref{obs} we discuss advantages and disadvantages of different types of magnetogram data. Section~\ref{anal} contains a discussion of the conditions in the photospheric magnetic field and how this leads naturally to the algorithms presented in Section~\ref{res}. The algorithms of turbulence and fractals described in this work are based on original work by Kolmogorov (1941) and Mandelbrot (1983) and have since been found to ubiquitous in many areas of human and natural sciences  (e.g., heartbeat dynamics, hydrology). These represent novel approaches to analyzing longitudinal magnetogram data and aim to generate physically motivated complexity measures. We conclude by discussing some of the drawbacks of these methods and possibilities for the future in Section~\ref{dis}.

\section{Data}
\label{obs}

In an ideal world, we could measure the full in-situ magnetic field vector from the photosphere to the corona, from which we could directly calculate any parameter of magnetic energy.. However, even in this seemingly perfect experiment, we would undoubtedly not find a perfect prediction of solar flare activity  - this arises dues to the inherent non-linear nature of the solution of the Navier-Stokes equation, and dependence on small errors in the initial conditions. In our real world, we are limited to remote sensing of the magnetic field at one or two heights in the atmosphere, which brings about a number of extra issues of the resolution and coverage in the spectral, temporal and spatial domains (McAteer et al. 2005b). Spectrally, we would like to measure the full vector field at multiple heights throughout the solar atmosphere. Temporally, we would like a full solar cycle of data from one instrument at a cadence higher than the evolution timescale of energy build-up. Spatially we would like to resolve the smallest features (i.e., elementary flux ropes), while covering the entire solar disk. Anything less than this ideal dataset and we have to be very careful of selection effects. Current instrumentation allows us to choose between full-disc longitudinal-measurements over a long timerange or partial disc vector field measurements of a few dozen regions. Specifically, the Michelson Doppler Imager (MDI; Scherrer et al. 1995) provides 96-min cadence, $1.96"$/pixel, full disc images of the photospheric longitudinal (from Stokes V and I) field and has operated since 1997. Complementary to this, the Global Oscillations Network Group (GONG; Harvey et al. 1996) also provides full disc photospheric longitudinal magnetograms since 1995 and the Synoptic Optical Long-term Investigations of the Sun  (SOLIS; Keller et al. 2001) provides similar data for the chromosphere. Small field-of-view vector field measurements are available from the Imaging Vector Magnetograph (IVM; Mickey et al. 1996) and Hinode Solar Optical Telescope (SOT; Tsuneta et al 2008). Soon we expect full disc vector magnetic field images from SOLIS, the COronal Solar Magnetism Observatory (COSMO) and the Helioseismic and Magnetic Imager (HMI). 

To add to this quagmire of confusion, the choices of data and algorithms are intrinsically interlinked. For space weather purposes the data must be reliable, readily available, and easy to obtain. Several measurements of helicity and non-potentiality are tied to vector data and so we must wait on HMI before we can fully exploit their potential and then we must wait several years before having enough data to test any flare prediction system. In the meantime our algorithm of choice must be appropriate for the longitudinal data (e.g., MDI) while producing a physically meaningful measure of flare production. Furthermore, it must run efficiently and quickly. As the algorithms we discuss in Section~\ref{res} are naturally scale free, as the data are available in near-realtime, and as we find that a full-solar cycle of regular data is an essential aspect, we suggest MDI will provide us with the best data for flare prediction over the next few years. 
	
\section{The Magnetic Reynolds Number}
\label{anal}

It seems that the magnetic field in active regions is the only viable means of storing and releasing the energy to drive solar flares. A simple discussion of the equations used to describe this magnetic field leads to a couple of possible choices of characterising the field. 

The induction equation is given by eliminating the electric field between Faraday's law and Ohm's law
\begin{equation}
\label{e1}
{%
\frac{\partial {\bf B}} {\partial t} = \nabla \times ( {\bf v} \times {\bf B} ) - \nabla \frac{j}{\sigma} ,
}
\end{equation}
which further reduced by Ampere's Law to
\begin{equation}
\label{e2}
{%
\frac{\partial {\bf B}} {\partial t} = \nabla \times( {\bf v} \times {\bf B} ) - \nabla \times ( \eta\nabla \times {\bf B}) ,
}
\end{equation}
where $\eta  = 1/ \mu_{0} \sigma$ is the magnetic diffusivity. By Gauss' law, and by means of a simple vector identity (and assuming constant $\eta$), this further reduces to the more familiar form of the induction equation,
\begin{equation}
\label{e3}
{%
\frac{\partial {\bf B}} {\partial t} = \nabla \times ( {\bf v} \times {\bf B} ) + \eta \nabla^{2} {\bf B} .
}
\end{equation}
This states that any local change in the magnetic field is due to a combination of a convection and a diffusive term. The ratio of these two terms is described by the magnetic Reynolds number,
\begin{equation}
\label{e4}
{%
R_{m}= \frac{\nabla \times ( {\bf v} \times {\bf B} )} { \eta \nabla^{2} {\bf B}} \approx \frac{\upsilon l }{\eta}
}
\end{equation}
which acts as a indication of the coupling between the plasma flow and the magnetic field. For typical photospheric values ($l \approx 10^{5}Mm,  {\bf v} \approx 10 ms^{-1}, \eta \approx 10^{3} m^2 s^{-1}$), the magnetic Reynolds number is much greater than one.  In this $R_{m} \gg 1$ regime, flux lines of the magnetic field are advected with the plasma flow, until such time that gradients are concentrated into short enough length scale that diffusion can balance convection ($l \approx 100m$), Ohmic dissipation becomes important, and magnetic reconnection can occur. Essentially the large $R_{m}$ allows for the build up of energy, followed by a sudden release. Physical system with large $R_{m}$ naturally lead to studies of {\em turbulence} and {\em complexity} as key parameters to developing a deeper understanding of the physics behind active region evolution and flare production.

\section{Turbulence and Complexity Measures}
\label{res}

The high magnetic Reynolds number in the photosphere suggests two possible, interlinked, methods of quantifying complexity. These are studies of the scaling index (arising from fully-developed turbulence) and fractality (arising from self-similarity). Previous studies have shown that the local properties of the active region field - critical in many theories of activity - are lost in the common global definition of their diagnostics, in effect smoothing out variations that occur on small spatial scales. While traditional Fourier and fractal methods have been used extensively in image processing, and are well understood, they do not provide a complete diagnostic of the range of spatial frequencies or scales within an image. Hence measures that are sensitive to the small-scale nature of energy storage and release in the solar atmosphere are required. As flows on the Sun exist in a state of fully developed turbulence, multiscale methods may be key to measuring and understanding the complex structures observed on the solar surface. In a similar fashion, multifractal, rather than monofractal, studies may be necessary to study the full complexity in locally inhomogeneous data. Importantly, these parameters can be related directly to predictions from theories of turbulence (Lawrence et al. 1993)

In this work we describe two complimentary techniques of multiscalar and multifractal methods. Both methods are based on the premise that physical systems cannot be adequately described by simple parameters, but require methods that capture their true complexity as a function of size scale. 

\subsection {Multifractal measures} 
The fractal dimension of any object can be thought of as the self-similarity of an image 
across all scale sizes, or the scaling index of any length to area measure, 
\begin{equation}
\label{e5}
{%
A \propto l^{\alpha}, 
}
\end{equation}
where $\alpha$ is the singularity strength. However, a multifractal system will contain a spectrum of singularity strengths of different powers, 
\begin{equation}
\label{e6}
{%
A \propto l^{f(\alpha)},
}
\end{equation}
and takes account of the measure at each point in space. Any measure distribution (e.g. magnetic field in an image) can be characterized by 
\begin{equation}
\label{e7}
{%
\psi (q,\tau) = E  \sum_{N}^{i=1} P^{q}_{i} \epsilon^{-\tau} 
}   
\end{equation}
where $q$, $\tau$ can be any real numbers, and E is the expectation of the object consisting of N parts. In this form, $\psi$ is the coupled $\tau$ -moment of the size $\epsilon$, and $q$-moment of the measure $P$. The three main multifractal indices commonly used to represent a non-uniform measure are then the : \newline
(i) Generalised Correlation dimensions (Grassberger \& Procaccia 1983), $D_q = {\tau}/(q-1)$; \newline
(ii) singularity strength, ${\alpha} = d{\tau}/dq$; \newline
(iii) Legendre transformed $f({\alpha}) = q{\alpha}-{\tau}$.\newline 
When applied to a traditional box-counting approach, it is useful to define the partition function, 
\begin{equation}
\label{e8}
{%
Z_{q}({\epsilon})=\sum_{i} P^{q}_{i}({\epsilon}),
}
\end{equation}
such that ${\tau} (q) = lim_{{\epsilon} {\rightarrow} 0} log (Z) / log ({\epsilon})$, and any of the three representations above can be calculated. The terminology used in this description is deliberately similar to that use in turbulence studies (McAteer et al. 2007). The $q$ moment plays the role of increasing the relative importance of the more intense parts of the measure as $q$ is increased. In this way it acts as a microscope to investigate the different contributions made to the image at higher values of the measure.

Another method of calculating the multifractal spectrum is based on the structure function (Parisi \& Frisch 1985; Abramenko et al. 2002). This consists of calculating the statistical moments of the field increments $S_q(r)$, as a function of separation, $r$ in order to determine the scaling exponents, $\zeta_q$
\begin{equation}
\label{e_str}
{%
S_q(r) \sim r^{\zeta_q}
}
\end{equation}
which are also directly related to the $f(\alpha)$ spectrum as (Muzy, Bacry \& Arneodo 1993)
\begin{equation}
\label{e_zeta}
{%
D(h) = qh - \zeta_q +1
}
\end{equation}
This formulation also provides a direct link to the Fourier scaling index (Section~\ref{multiscale}) as $\beta \equiv 1 - \zeta_6$

Another recent advance in this field uses the properties of the wavelet transform to order to overcome the limitations of the box-counting method (Conlon et al. 2009, Kestener et al. 2009). In this terminology two properties,$D(h_q)$ and $h_q$, are calculated separately by chaining the modulus maxima of the wavelet transform of an image, resulting in a $D(h)$ spectrum which is directly related to the $f(\alpha)$ spectrum as \newline
(i) ${f(\alpha)} = D(h_q)$,\newline
(ii) $h_q = \alpha - E_{dim}$,\newline
where $E_{dim}$ is the euclidian dimension ($E_{dim} = 2$ for an image). This wavelet transform modulus maxima (WTMM) formulation offers many computational advantages over the box-counting and structure function approaches, including more accurate results at negative $q$ and large positive $q$.

\begin{figure}
\label{fig1}
\begin{center}
\includegraphics*[width=12cm]{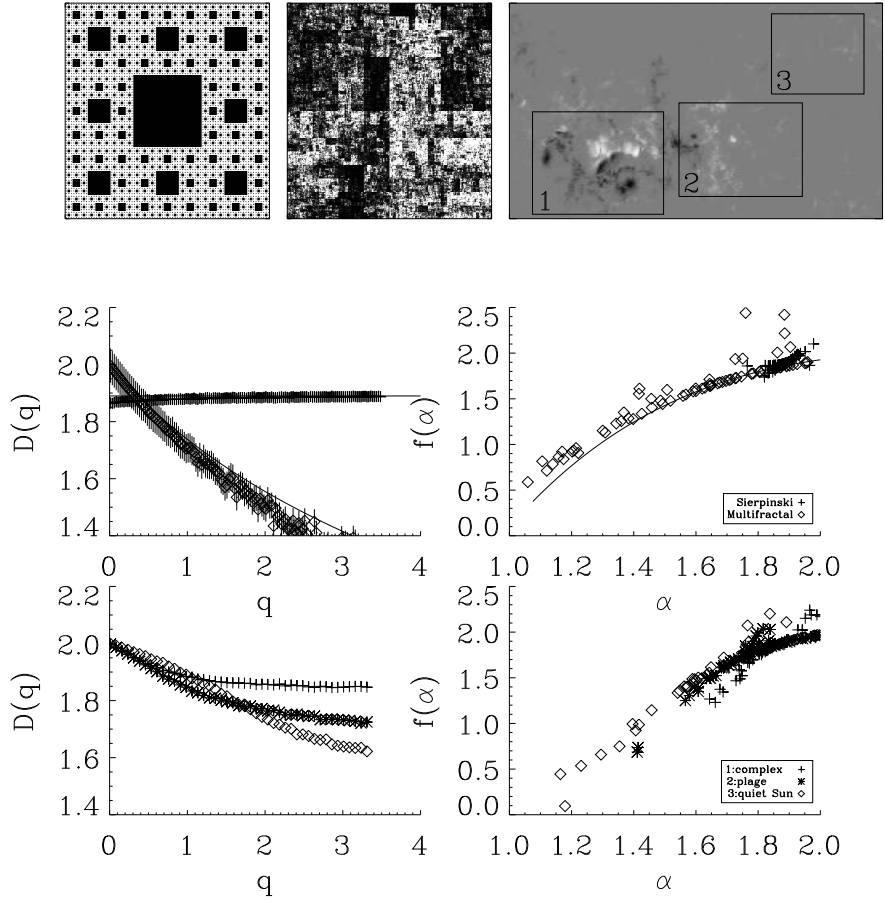}
\end{center}
\caption{Top: An example of a monofractal (Sierpinski carpet; left), a multifractal (middle),  and an active region (NOAA 9077) from MDI (right). Middle: The $D_q$ and $f(\alpha)$ spectrum of the monofractal (crosses) and multifractal (diamonds). The solid lines correspond to the theoretical values. Bottom: The $D_q$ and $f(\alpha)$ spectrum of three segments of the image of the active region}
\end{figure}

Figure 1 shows a comparison of the $D_q$ and $f(\alpha)$ spectra for a monofractal (top left; the Sierpinski carpet), a multifractal (middle), and a solar active region magnetogram (right). The monofractal is described as a flat $D_q$ spectrum ($D_q \sim 1.89$ for all q)- it demonstrates the same complexity at all moments -  and consequently a narrow $f(\alpha)$ spectrum. In the limit of an ideal algorithm, $f(\alpha)$ would consist of a single point at $f(\alpha=2) = 2$. The multifractal exhibits a monotonically decreasing $D_q$ spectrum, which can be interpreted as meaning the image consists of series of fractals, each of which dominate at different moments. The $f(\alpha)$ spectrum is wide and drops off rapidly which shows that the image contains a large number of singularity strengths each with a different power. Multifractals always show a decreasing $D_q$ and more complex images contains more power at large singularity strengths. It is important to note that the degree of mutlifractality (e.g., the width and height of $f(\alpha)$) reflects the range of fractals which exist in the image.

The three segments of the MDI image all display a distinct drop off with increasing $q$, and hence are all multifractal. The complex segment shows a high $D_q \sim 1.85$ at large $q$, and a narrow, well-peaked $f(\alpha)$ spectrum. This in is interpreted as meaning that the regions of strong field within this segment are highly complex, and contribute as much complexity to the image as the lower magnetic field parts. This can be contrasted with the plage segment of the data, which exhibits a $D_q$ spectrum with a similar drop off at small $q$ but continues to drop off at larger $q$. Correspondingly, the  $f(\alpha)$ spectrum is much wider. Hence the strong field parts of the plage segment are rare and less complex. Finally, the quiet-Sun segment displays a $D_q$ spectrum which is much lower at large $q$. The $f(\alpha)$ is very sparse at the high $\alpha$ values and drops off to very low $f(\alpha)$ at the lower $\alpha$ values. Hence the quiet Sun segment is mostly small, weak field. 

We can compare the differences in these spectra to the probabilities of flares occurring in each part of the image. Multifractal measures can be directly compared to self-similar cascade or self-organized criticality models, both of which have been used to model energy release in solar flares (Lawrence et al. 1993; Georgoulis et al 2002). Fractal and multifractal algorithms have been applied extensively to photospheric magnetic field data. Abramenko et al. (2002) found that the relative fraction of small scale fluctuation in the magnetic field contribute significantly more prior to flaring. Abramenko et al. (2005a) found that active regions reach a critical state of intermittency prior to flaring. McAteer et al. (2005a) tested a simple monofractal approach on the largest dataset to date, and found a minimum $D_q =1.2$ is a necessary, but not sufficient requirement for M and X-class flares. Conlon et al. (2008, 2009) find two distinct thresholds in the multifractal spectrum which correspond to the onset of flare production in an otherwise quiet active region. They confirm a minimum Haussdorf dimension of $D(h) =1.2$  is a necessary, but not sufficient requirement for M and X-class flares. Furthermore this must be accompanied by a minimum H\"{o}lder exponent of -$0.7$. Physically this corresponds to a global restructuring of of the field distribution from Dirac like noise ($h = -1$) to a step-function ($h = 0$) form. As the H\"{o}lder exponent increases towards zero, salt-and-pepper type noise is replaced by the formation of gradients, permitting the build up of energy in the system
			
\subsection {Multiscalar measures} 
\label{multiscale}

It is recognised that small regions of flux emergence / cancellation are vital in detailing the evolution of solar active regions. For this reason the wavelet transform can be adopted as it localised in space and hence allows for the detection of local image 
features. 

The continuous wavelet transform of an image, $I(r)$ can be defined as 
\begin{equation}
\label{e9}
{%
w (s, x) = \frac{1}{\sqrt{s}} \int^{\infty}_{-\infty} I(r) \Psi^{*} \left(  \frac{r - x}{ s } \right) d^{2}r,
}
\end{equation}
where $\Psi$ is the mother wavelet, $s$ is a term describing scale at a position $r$, and $w(s,x)$ are the 
wavelet coefficients of the image. The mother wavelet can take several forms, depending on the application. Wavelet analysis retains the localized spatial information, providing vital information on the turbulent flow. As such, a wavelet analysis provides an indispensable complement to a multifractal analysis.  The choice of mother wavelet is determined by the science requirements. A `derivative of Gaussian' (also know as mexican hat) wavelet is useful for detecting sudden local changes in an image hence this can be use to detect the neutral line position and magnitude. A `Haar' wavelet is useful in detecting and removing the intermittent component of an image (McAteer \& Bloomfield 2009), a vital tool in describing bursty behaviour. A discrete (e.g., `\`{a} trous') wavelet can be used to efficiently detect flux emergence as a function of size scale.  

\begin{figure}
\label{figure2}
\begin{center}
\includegraphics*[width=12cm]{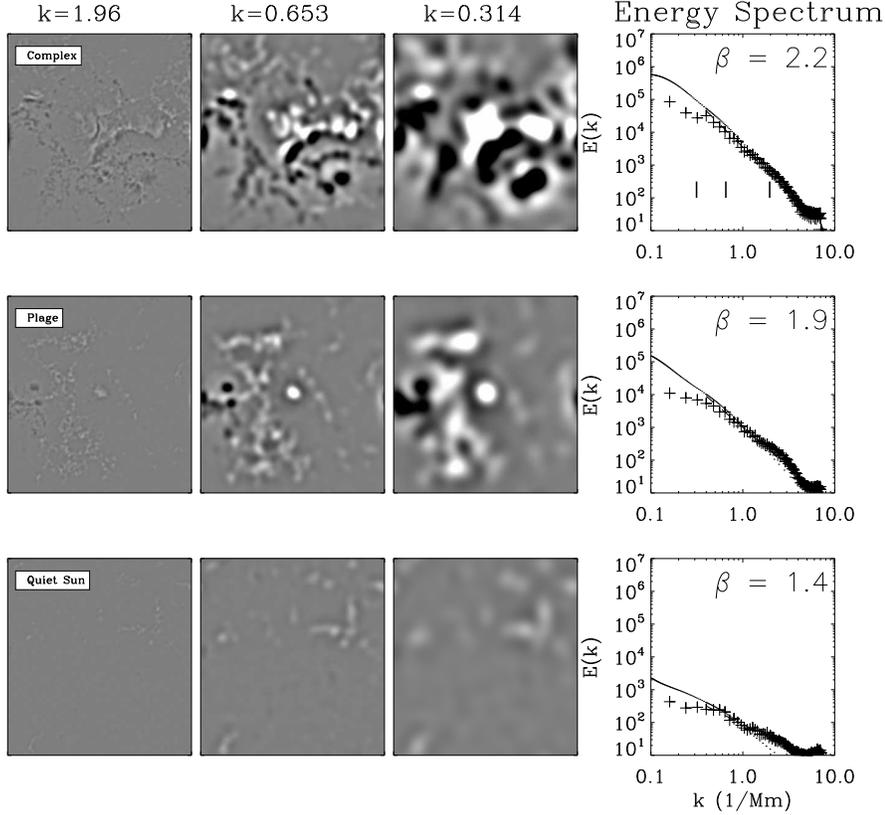}
\end{center}
\caption{A wavelet decomposition of the complex segment (region 1; top),plage segment (region 2: middle), and quiet Sun segment (region 3: bottom) of the active region from Figure 1 at scales of 3.2~Mm (1st column), $9.6$~ Mm (2nd column), and $20.0$~Mm (third column). The fourth column shows the energy spectrum as calculated from the Fourier technique (crosses) and the wavelet analysis (dots)}
\end{figure}
	
Figure 2 shows the wavelet analysis decomposition (using the mexican hat wavelet) at three spatial scales for the same three segments of the MDI data as in Figure 1. The nine decompositions are all plotted in the same grayscale range. This makes it clear, that the energy (which is just the sum of the square of each value in the image)
drops off from large k to small k for any image (reading across the rows). At any one scale, the energy is higher in the complex segment than in the plage, and is higher in the plage than in the quiet Sun (reading down the columns). The final column shows the energy spectrum as calculated using both the normal Fourier technique (adopted from Abramenko et al. (2005b)), and the wavelet technique (adapted from Hewett et al, (2008)). The calculation of the energy spectrum helps to characterize the intermittency of an image. Kolmogorov (1941) showed that the energy spectrum ($E(k)$ of a system, scales as the wavenumber, $k$,
\begin{equation}
\label{eq_kol}
{%
E(k) \sim k ^{-\beta}
}
\end{equation} 
where $\beta = 5/3$ for fully developed turbulence. In Figure 2, $\beta$ is calculated from a linear regression of the $E(k)$ plot over the 3-10Mm range (as suggested by Abramenko et al. (2005b)). It is clear that the complex segment contains more power than the plage (and quiet Sun ) at all $k$ and preferentially contains much more power at small $k$ (hence large spatial scales) which results in a larger scaling index. Hence a large $\beta$  is suggestive of increased complexity and more, larger, solar flares as demonstrated in Abramenko et al. (2005b).

There are two main advantages of the wavelet technique over the Fourier technique, both of which are clear from Figure~2. Firstly, the linear range of the energy spectrum is much larger and hence the value of $\beta$ calculated is much less dependent on identifying the correct range of linearity. The Fourier technique must exhibit a drop off for $E(k)$ for small $k$ (where the spatial scale is too large compared to the image size) and large $k$ (where the spatial scale is not periodic across the image). The inherent ability of the wavelet analysis to adapt to both small and large $k$ makes the resulting energy spectrum more linear. Hewett et al. (2008) showed that the energy spectrum of an image can be accurately calculated over a larger range of scales using a wavelet transform. This work found a sudden onset of flares from an otherwise flare-quiet active region when the $\beta$ suddenly dips from a previous value of $-1$ to a much steeper value of $-3$. The flare onset seems to occur when the spectrum passes thru the Kolmogorov index of $-5/3$.  Furthermore the linear range of the scaling increases a few days prior to a flare. The linear range is theoretically known to increase with increasing $R_m$ but the exact relationship is still under study. 

The second advantage is the retention of the spatial information in the the decomposition from which we can extract physical parameters at different scales. At the smallest spatial scale, there is a clear neutral line near the center of the complex image. However at the largest scale, the neutral line towards the lower right of the image is more significant.  Ireland et al. (2008) use the wavelet transform to decompose magnetic field images into different scale sizes and show these are related to different Mount Wilson classes. They also find a significant difference in the distribution of gradients between flaring and non flaring regions and find this is maintained over all scales. The wavelet transform helps to encode the ideas of magnetic gradients and flux emergence over all length scales, not just those where the strongest field are found.
	
\section{Discussion}
\label{dis}

We have discussed the necessity for accurate solar flare prediction and have reviewed the significant progress made over the past few years using parameters of turbulence and complexity. It is clear that the more complex regions produce more larger flares, and by using multiscale and multifractal techniques we may be able to capture this notion of complexity into one or two parameters. These take advantage of the excellent availability of a wide range of solar data both present and expected in the future. Data from new instruments will undoubtedly advance our understanding of active regions and the mechanisms through which they may affect human society. These new data will require new algorithms, e.g., calculating the multifractal spectrum of a vector field, which are currently being developed. This is expected to provide significant progress in predicting the true flare potential of individual active regions in near-realtime.

There are a number of obvious problems with these techniques /data which must be addressed over the next few years. Firstly, we are only measuring the driving component of the build-up of energy - this does not inevitably lead to solar flares. The mechanism to initiate a flare can probably only be identified by studying the transverse component of the magnetic field in the corona. Secondly we are not fully aware of the natural timescale of the Sun. Although humans may prefer a 24-hour prediction, the Sun has a multitude of differing timescales interacting. It appears the driver of energy build-up may occur over days,  with sudden inputs over hours being most significant. There may be a strong solar cycle dependence - predictive tools applicable to cycle 23 may not work for cycle 24. Furthermore flares may not be Poisson distributed - it is well known that one of the best predictive signs of a large flare is a previous large flare. Thirdly we are not yet fully aware of what we mean by a {\em large} flare. We are only beginning to study the energy distribution  (between radiation, particles, and the coronal mass ejection) in a flare event (Emslie et al. 2004). For space weather purposes, the geo-effective quality is a much more important statistic and this does not necessarily scale linearly with the scientists' measure (e.g., GOES class). We also have to be careful not to only study large events. A full physical understanding requires our models to work over all flare sizes. Lastly all algorithms need to be fully tested using climatological skill scores in order to make the step from scientifically interesting to space-weather prediction (Bloomfield et al. 2009)

The author wishes to thank the organisers of the sessions at European Space Weather 2008 for the invitation to present this work. This paper was greatly enhanced by discussions at the NASA all-clear workshop with Barnes, Leka, and Gourgoulis. The content of this paper was significantly improved by the comments of two anonymous referees. The authors thank the SOHO/MDI consortia for making data available. This research was supported by a grant from the Ulysses Ireland-France Exchange Scheme operated by the Royal Irish Academy and the Ministre des Affaires Etrangres. Paul A. Conlon is an IRCSET  Government of Ireland Scholar. R.T.James McAteer is the recipient of a Marie Curie Intra-European Fellowship under FP6.

\clearpage

\end{document}